\documentstyle[aps,prl,twocolumn,epsf]{revtex}
\begin{document}
\input{psfig.sty}
\draft

\newcommand{\bQ}{\mbox{${\bf Q}$}}
\newcommand{\h}{\mbox{$\frac{1}{2}$}}
\newcommand{\th}{\mbox{$\frac{3}{2}$}}
\def\bolda{\mbox{\boldmath$a$}}
\def\boldb{\mbox{\boldmath$b$}}
\def\boldc{\mbox{\boldmath$c$}}
\def\boldQ{\mbox{\boldmath$Q$}}
\def\boldq{\mbox{\boldmath$q$}}
\def\btau{\mbox{\boldmath$\tau$}}

\twocolumn[\hsize\textwidth\columnwidth\hsize\csname
@twocolumnfalse\endcsname

\title{Continuum in the spin excitation spectrum of a Haldane
chain, observed by neutron scattering in CsNiCl$_3$. }

\author{I.~A.~Zaliznyak$^{1}$, S.-H.~Lee$^{2,3}$, S.~V.~Petrov$^4$}
\address{
 $^1$Department of Physics, Brookhaven National Laboratory, Upton,
 New York 11973-5000 \\
 $^2$National Institute of Standards and Technology, Gaithersburg,
 Maryland 20899\\
 $^3$Department of Physics, University of Maryland, College Park,
 Maryland 20742\\
 $^4$P.~Kapitza Institute for Physical Problems, ul. Kosygina, 2,
 Moscow, 117334 Russia
 }

\date{\today}
\maketitle

\begin{abstract}

The spin excitation continuum, expected to dominate the low-energy
fluctuation spectrum in the Haldane spin chain around the
Brillouin zone center, $q=0$, is directly observed by inelastic
magnetic neutron scattering in the $S=1$ quasi-1D antiferromagnet
CsNiCl$_3$. We find that the single mode approximation fails, and
that a finite energy width appears in the dynamic correlation
function ${\cal S}(q,\omega)$ for $q\lesssim 0.5\pi$. The width
increases with decreasing $q$, while ${\cal S}(q,\omega)$ acquires
an asymmetric shape qualitatively similar to that predicted for
the 2-magnon continuum in the nonlinear $\sigma$-model.

\end{abstract}

]

Experiments on low-dimensional magnets are an invaluable source of
insight into the fundamental properties of quantum spin models and
field theories, many of which are still awaiting theoretical
solutions. A vivid example is provided by the one-dimensional (1D)
$S=1$ Heisenberg antiferromagnet (HAFM). It was conjectured by
Haldane \cite{Haldane}, and then confirmed by experiments
\cite{Regnault,Broholm,Buyers,Steiner} and numerical studies
\cite{Takahashi,Meshkov,Yamamoto,ED,White,Sorensen}, that its
ground state (GS) exhibits a robust quantum disorder,
characteristic of the 1D nonlinear $\sigma$ model (NL$\sigma$M).
In striking contrast with the almost ordered "marginal liquid"
state of the $S=1/2$ chain \cite{Bethe}, the Haldane GS is a
"dense spin gas", with a short correlation length of $\xi\approx
6$ lattice repeats, and a gap $\Delta_H \approx 0.41J$ ($J$ is the
exchange coupling) in the spin excitation spectrum. The
excitations' spectral weight is concentrated  in a long-lived
massive triplet mode in the vicinity of the Brillouin zone (BZ)
boundary $q=\pi$, somewhat similar to the spin-wave doublet in the
classical S$\gg 1$ AFM with easy-axis anisotropy \cite{Buyers}.
Any remains of the spectacular continuum at $q\approx\pi$, found
in the $S=1/2$ HAFM chain \cite{Bethe,Dender}, is attributed to
3(5,7,...)-magnon states, and predicted to be extremely weak
\cite{Takahashi,Meshkov,Horton}. Although no quantitative theory
of the excitation spectrum in the whole BZ exists, a continuum is
expected in the vicinity of the zone center $q=0$, where the
lowest energy excitations are pairs of $q\approx \pi$ magnons.

The two-magnon nature of the lowest excited state in the 1D $S=1$
HAFM at $q\approx 0$ was first stipulated by the Monte Carlo (MC)
numerical study \cite{Takahashi}. It established, that the gap in
the magnon spectrum at $q=0$ is about twice as large as that at
$q=\pi$. It was later supported by MC simulations on larger
systems \cite{Takahashi,Meshkov,Yamamoto}, and exact
diagonalization (ED) for rings with up to 18 spins
\cite{Takahashi,ED}. A semi-quantitative illustration of the
lattice periodic quasiparticle dispersion, crossing the
two-particle energy at a $q\lesssim \pi/2$, was obtained in the
physically motivated variational theory \cite{Gomez-Santos}, based
on the mapping of the spin Hamiltonian on a system of spinless
fermions. Not only does this simple theory correctly capture the
fermion-like nature of the $q\approx\pi$ magnons observed in
experiment \cite{Zaliznyak1993}, but it also gives a correct
functional form for the single-magnon dispersion,

\begin{equation}
\label{Eq}
 \varepsilon(q)= \sqrt{\Delta_H^2+v^2\sin^2{q}+\alpha^2\cos^2 \frac{q}{2}}\;,
\end{equation}
which for $q\gtrsim\pi/2$ is in a very good agreement with MC
\cite{Takahashi}. In fact, the dispersion, measured by neutron
scattering in the Haldane chain compound NENP \cite{Broholm}, was
found to coincide with (\ref{Eq}) in a large part of the BZ, at
$q\gtrsim 0.3\pi$, if the parameter $\alpha$ is adjusted from
$\alpha\approx 2.5J$ \cite{Gomez-Santos} to $\alpha\approx 1.45J$.
In this paper we use (\ref{Eq}) to describe the 1D excitation
spectrum in CsNiCl$_3$ (solid lines in Fig.1,3). We find that it
gives an excellent fit to the measured excitation energies,
bringing an end to the existing controversy \cite{Buyers}. Using
$J=2.275$ meV, independently determined from the high field
magnetization \cite{Katori}, we obtain $v=2.49(4)J$, in a
remarkable agreement with the numerical result \cite{Sorensen}.
Most importantly, we find that spectral density deviates from the
SMA at $q\lesssim 0.5\pi$, and present the first resolved
measurement of the continuum part of the excitation spectrum of a
Haldane spin chain.

In spite of extensive theoretical evidence for existence of the
$q\approx 0$ continuum in the spectrum of a 1D $S=1$ HAFM, its
limits and extent remain unclear and controversial, while its
experimental characterization is considered practically impossible
because of the rapid decrease of the static spin structure factor
${\cal S}(q) = \int {\cal S}(q,\omega) d(\hbar \omega)$ at small
$q$. This follows from the first moment sum rule
\cite{HohenbergBrinkman}, which establishes an {\it exact}
relation between ${\cal S}(q)$ and the excitation average energy
$\langle\varepsilon(q) \rangle = \int (\hbar \omega) {\cal
S}(q,\omega) d(\hbar \omega) /\int {\cal S}(q,\omega) d(\hbar
\omega)$. For the 1D HAFM,

\begin{equation}
\label{Sq1D}
 {\cal S}(q)= - \frac{2}{3} E_{GS} \frac{(1-\cos{q})}{\langle \varepsilon(q)
\rangle} \;,
\end{equation}
where $E_{GS}=-1.40(0)J$ \cite{White} is the GS energy per site.
For a gapful spectrum ${\cal S}(q)$ of (\ref{Sq1D}) vanishes $\sim
q^2$ at $q\rightarrow 0$. Where the SMA holds, ${\cal S}(q)$ is
uniquely determined by the dispersion, $\langle \varepsilon (q)
\rangle \equiv \varepsilon(q)$. In addition, experimental
characterization of the continuum is challenged if the excitation
spectrum is split by the anisotropy, as in NENP and related
Ni-organic model compounds, or if the excitations acquire a
temperature-activated damping.

CsNiCl$_3$ is one of the most isotropic and best studied quasi-1D
$S=1$ HAFM model compounds (see \cite{Buyers,Steiner} and
references therein). It has a hexagonal crystal structure, space
group P63/mmc; at $T=1.5$ K the lattice spacings are $a=7.12$\AA,
$c=5.9$\AA. Chains of Ni$^{2+}$ ions run along the $c$ axis and
form a triangular lattice in the $a-b$ plane. There are two
equivalent ions per $c$ spacing, so that $\boldQ=(h,k,l)$ in
reciprocal lattice units (rlu) corresponds to $q = \pi l $ in the
1D BZ of a chain. A very reliable estimate for the in-chain
exchange coupling, $J=2.275$ meV, is obtained from the measured
spin-flip (saturation) field $H_s=4Jg\mu_BS = 73.5$ T
\cite{Katori}. A spin-flop (reorientation) field $H_{sf} \approx
1.9$ T implies a negligible single-ion anisotropy, $D\approx
0.002J$.

We measured the dependence of the energy spectrum of the spin
dynamic structure factor on the wavevector transfer along the
chains for a CsNiCl$_3$ sample made of two large crystals with
total mass $6.4(1)$ g, at $T=1.5$ K. The sample was mounted with
the $(h,h,l)$ zone in the scattering plane, and had effective
mosaic spread $<1^\circ$. To increase the data collection rate,
the SPINS 3-axis cold neutron spectrometer at NIST Center for
Neutron Research was equipped with a position sensitive detector
(PSD), matched in size to the large PG(002) flat analysing
crystal. The analyser's central energy was fixed at
$E_f^{(0)}=4.2$ meV; its angular acceptance was $\approx 9^\circ$.
Calibration of the neutron final energy and the sensitivity across
the PSD was done using elastic incoherent scattering from
Vanadium. Beam divergence was defined by the $^{58}$Ni neutron
guide and a $80'$ radial collimator in front of the PSD. To
perform energy scans at constant $q=\pi l$, the wavevector
transfer along $[110]$ was varied, so that chains were always
aligned with the analyser [002] Bragg wavevector $\btau_A$.
Typical variation of $h$, imposed by this condition, is
illustrated by the right and top axes of Fig. 1, which shows the
contour plot of the raw spectral density of the scattering
intensity ($\sim I(q,E)/\int I(q,E)dE$), with the $q$-dependent
flat background subtracted. It is evident from Fig. 1, that the
spectrum acquires a finite width in energy at $l \lesssim 0.5$.
Although the effect is somewhat exaggerated by the resolution,
careful accounting for the latter shows that {\it intrinsic}
width, where it is non-zero, accounts for $\sim 2/3$ of the total
width. Curves show the prediction of \cite{Gomez-Santos}, obtained
from (\ref{Eq}) with $J=2.275$ meV, $\Delta_H=0.41J \;, v = 2.49
J$, and $\alpha = v$.

For the quantitative analysis of the energy dependence of the
measured cross-section, we use the normalized scattering function
of the damped harmonic oscillator (DHO), parametrized in terms of
the position $\omega_0$, and the full width at half maximum (FWHM)
of the corresponding antisymmetrized Lorentzian peak
\cite{Steiner}. Alternatively, to describe the asymmetric peak
shape at lower $q$, we use a "half-Lorentzian" truncated DHO
(TDHO), which is obtained by multiplying the above by
$\theta(\omega^2 - \omega_0^2 )$. The TDHO fits to several
constant-$q$ scans are shown in Fig.~2. As quantified by $\chi^2$
in Fig. 3(d), TDHO is in better agreement with experiment at
$q\lesssim 0.5\pi$, where a sharper onset of the scattering at low
energies, characteristic of a continuum, is observed. Note, that
an {\it opposite} asymmetry \cite{Broholm} would result from the
interplay of the dispersion with the instrument resolution.

\begin{figure}[t] \noindent\vspace{0.1in}
\parbox[b]{3.4in}{
\psfig{file=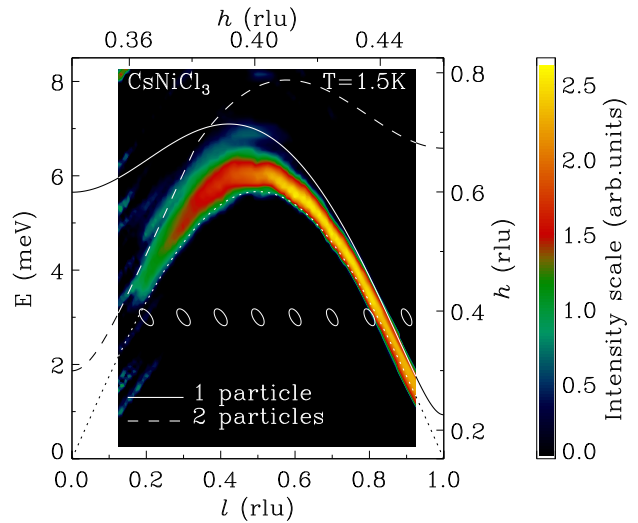,height=3.5in} \vspace{-1.in} \noindent
\caption{Contour plot of the measured spectral density of magnetic
scattering, reconstructed from 9 constant-$q$ scans via linear
interpolation. Scale on the right shows variation with energy of
the wavevector transfer perpendicular to the chain at $l=0.5$,
scale on the top -- its variation with $l$ at $E=3$~meV. Ellipses
are the half maximum contours of the instrument resolution
function, calculated at $E=3$~meV. Solid curve is the
single-magnon dispersion (\ref{Eq}), dashed line shows the lowest
energy of the two non-interacting magnons with given total $q=\pi
l$, dotted line is $\varepsilon(q)= v\sin{q}$. } }
\end{figure} \vspace{-0.1in}

Peak parameters, refined for both DHO and TDHO fits, are detailed
in Fig. 3. The dispersion of the center of mass of the excitation
spectrum, captured by the position of the DHO peak (circles in Fig
1(a)), is nicely described by (\ref{Eq}). For $J=2.275$ meV, the
best fit, shown in the figure, gives $\Delta_H = 0.34(6) J$, $v =
2.49(4) J$, and $\alpha = 1.1(4)$.
While at $q\leq 0.9\pi$ the spectrum is not very sensitive to
$\Delta_H$, the agreement of the magnon velocity $v$ with
calculations \cite{Sorensen} is impressive. The energy integrated
intensity of both DHO and TDHO peaks, Fig. 3(b), is in good
agreement with the sum rule (\ref{Sq1D}). A slight deviation,
concomitant with non-zero peak energy width (Fig. 3(c)), is due to
the extra intensity in the unphysical high-energy tail, and can be
completely removed by imposing an upper boundary on the continuum.

\begin{figure}[t] \noindent
\parbox[b]{3.4in}{\psfig{file=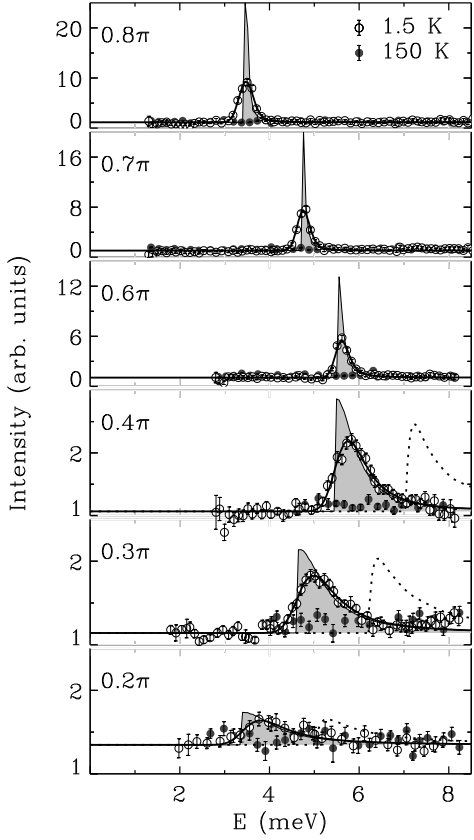,height=4.1in}
\vspace{0.45in} \noindent \caption{Energy spectrum of the
inelastic scattering, measured in CsNiCl$_3$, for several $q =
\boldQ\boldc/2 = \pi l$. The wavevector component in the $a-b$
plane $q_\perp = \boldQ\bolda + \boldQ\boldb = \pi h\sqrt{3}$ is
defined by the scattering geometry $\boldc\parallel\tau_A$ (Fig.
1). Solid circles show the non-magnetic background. Thick lines
are the resolution-corrected fits to the cross-section, based on
the TDHO scattering function. Shaded peaks are the same
cross-section, multiplied by the resolution volume, and illustrate
the "deconvoluted" intensity. Dotted lines show the cross-section
for the 2-magnon continuum in the NL$\sigma$-model
\protect\cite{AffleckWeston}, corrected for the non-relativistic
dispersion (\ref{Eq}). }}
\end{figure} \vspace{-0.1in}

The peaks at $q\lesssim 0.5\pi$, are very similar in shape and
intensity to the two-magnon continuum, calculated for the 1D
NL$\sigma$M \cite{AffleckWeston}, which correctly describes the
$S=1$ HAFM chain in the vicinity of $q = \pi$. We show the
NL$\sigma$M prediction by dotted lines in Fig 2.
Even though we used the realistic magnon dispersion (\ref{Eq}),
instead of the relativistic form $\varepsilon(q) =
\sqrt{\Delta_H^2+(v\tilde{q})^2}, \; \tilde{q}=q {\rm mod} \pi$,
which fails for $\tilde{q}\gtrsim 0.2\pi$, the calculated
continuum is still too high in energy. In fact, this discrepancy
does not simply show the limitation of the $\sigma$-model, but
also reveals the importance of magnon interactions. Indeed, a
careful examination of Figs. 1, 3(a) shows, that the lower
boundary of the observed continuum lies {\it below} the lowest
energy of two non-interacting magnons. This could be understood,
if there is an {\it attraction} between the quasiparticles, which
form the continuum states.

To understand how our findings compare with existing results, it
is important to realize that the extent of the observed continuum
is rather small, as it should be, according to the calculations
\cite{AffleckWeston}. At $q=0.3\pi$, the FWHM is $\approx 0.3 J$,
or only $\approx 15\%$ of the peak energy, and requires high
resolution and low temperature, $T\ll 0.3 J$, to be detected.
Previous attempts to measure the in-chain dispersion in CsNiCl$_3$
\cite{Buyers} were, on the contrary, done with rather coarse
resolution of the thermal neutron spectrometer, and at $T \sim 0.4
J$, where temperature damping dominates the spectral width. It is
more interesting to compare our data with the neutron scattering
study in NENP \cite{Broholm}, where the authors measured the
dispersion for $q\geq 0.3\pi$ at $T\lesssim 0.01J$, and found no
appreciable deviation from the SMA. Consequently, a picture for
the excitation spectrum of the 1D $S=1$ HAFM, where a single-mode
dispersion merges into a broad, but unmeasurable, continuum at $q<
0.3\pi$, became broadly popular \cite{Takahashi,White}. The
results of \cite{Broholm} are, however, easily reconciled with our
data, if we note that, to observe magnetic scattering at small
$q$, the authors had to relax the spectrometer resolution so much
that almost a quarter of the 1D dispersion was within their
wavevector acceptance. The effect we observe is simply unresolved
in such a measurement. The splitting of the spectrum by single-ion
anisotropy, which is rather large in NENP \cite{Zaliznyak1993},
presented another obstacle to the characterization of the
continuum.
A crossover to a continuum at $q \approx\pi/2$ is also supported
by numerical calculations \cite{Meshkov,Sorensen,Zaliznyak1993}.

Finally, the main disadvantage of CsNiCl$_3$ as a model 1D HAFM is
the supercritical inter-chain coupling $J_\perp$, which leads to a
3D order with the propagation vector $\boldQ_0 = (1/3,1/3,1)$, at
$T_N\approx 4.8$ K. The relevance of the Haldane conjecture in
this situation was the subject of a long-lasting controversy. It
was established that a gap opens in the spin excitation spectrum
at $T>T_N$, and in this so-called "1D phase" it recovers the
features of a Haldane chain. However, since $T_N\approx 0.2J$, the
spectra also acquire significant temperature broadening
\cite{Steiner}. In fact, we argue that 3D order in CsNiCl$_3$ is
so weak, that even at $T\approx 0$ it causes no significant change
in the spectrum throughout the better part of the BZ. This agrees
with the simple physical argument that excitations whose energies
are sufficiently large compared to $J_\perp$ are not sensitive to
it. Indeed, elastic Bragg intensity, corresponding to the ordered
spin value $\langle S\rangle = 0.5$, observed in experiment
\cite{Yelon}, accounts for only $\langle S\rangle^2 /(S(S+1)) =
12.5\%$ of the spin fluctuation spectrum. In the 1D $S=1$ HAFM
this fraction of the spectral weight is concentrated in the tiny
region of the BZ, at $|q| \gtrsim 0.98\pi$. Experiments also show,
that the Haldane gap triplet mode is split in the 3D ordered phase
only in the close vicinity of the magnetic Bragg peaks, at $|q|
\gtrsim 0.9\pi$, where the Goldstone acoustic magnons appear
\cite{Buyers,Steiner}.

To quantify our arguments, we consider corrections to the ${\cal
S}(\boldq)$ of (\ref{Sq1D}) for the ordered quasi-1D HAFM. The
right-hand side is modified in three ways. First, the exchange
energy of an individual chain changes on account

\begin{figure}[t] \noindent
\parbox[b]{3.4in}{\psfig{file=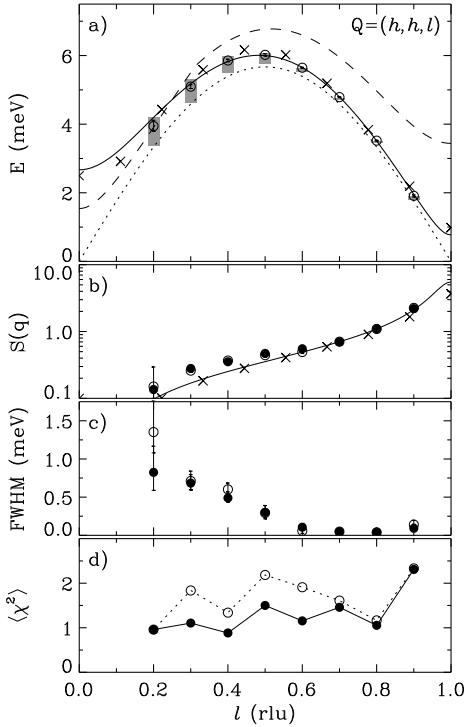,height=3.5in}
\vspace{0.35in}\noindent \caption{Wavevector dependence of the
excitation spectrum in CsNiCl$_3$. Full and open circles show the
parameters obtained for the asymmetric TDHO and symmetic DHO
lineshapes, respectively. Crosses are ED results \protect\cite{Takahashi}.
(a) Shaded bars start at the lower boundary of the asymmetric TDHO
peak, and extend by its FWHM. Circles show the Lorentzian peak
position of the symmetric DHO fits. Lines are the same as in Fig.
1, but with $\Delta_H=0.34 J ,\; \alpha=1.2 J$. (b) Static
structure factor. Line is the SMA result for the dispersion shown
in (a). (c) Peak (continuum) energy width. (d) Standard square
deviations for the TDHO and DHO fits.
 }}
\end{figure} \vspace{-0.1in}

\noindent  of the static order. In the mean field (MF)
approximation, at $\langle S\rangle \lesssim 0.5$, it increases by
$\langle S \rangle^2/(2\chi_\pi)$, which is less than $0.5\%$, due
to the large static staggered susceptibility of a Haldane chain
$\chi_\pi \approx 22/J$ \cite{ED,Zheludev}. Second, the numerator
has to be amended by adding the interchain correlation,
$\sum_{\beta,\btau_\perp} J_\perp (1 - \cos
(\boldq_\perp\cdot\btau_\perp)) \overline{\langle S_{\bf R}^\beta
S_{\bf R + \btau_\perp}^\beta \rangle}$, $\boldq_\perp$ being the
wavevector component, perpendicular to the chain. In MF this is
$\approx (J_\perp (0) - J_\perp(\boldq_\perp)) \langle S
\rangle^2/2 \leq (9/2)J_\perp \langle S \rangle^2$, where $J_\perp
(\boldq_\perp)$ is the Fourier transform of the interchain
coupling, and can be neglected for $\frac{9 }{2}J_\perp\langle S
\rangle^2\ll -\frac{2}{3} E_{GS} (1-\cos q) $. Finally, the
interchain dispersion appears, which changes the magnon average
energy in the denominator. In the random phase approximation
(RPA), $\varepsilon(q,\boldq_\perp) \approx \varepsilon(q)\sqrt{ 1
+ \chi_qJ_\perp (\boldq_\perp)}\approx \varepsilon(q) + {\cal
S}(q) J_\perp(\boldq_\perp)$, where $\varepsilon(q)$ and $\chi_q =
2{\cal S} (q)/\varepsilon (q)$ are the magnon energy and the spin
susceptibility for a single chain. From the boundaries of the
interchain dispersion at $q=\pi$, measured for CsNiCl$_3$ in
\cite{Buyers} at $T \approx 10$ K, and using $\chi_{\pi,T\approx
0.4J}\approx 0.74\chi_{\pi,T=0}$ \cite{Broholm}, we find
$max\{J_\perp (\boldq_\perp)\} - min\{J_\perp (\boldq_\perp)\} =
9J_\perp \approx 0.3J$. Plugging this value in the above
estimates, we conclude, that the purely 1D expression for the
static structure factor (\ref{Sq1D}), and, by definition, the
energy integrated scattering intensity, holds for CsNiCl$_3$
within $\pm 5\%$ at $0.3\pi\lesssim q\lesssim 0.7\pi$, and within
$\pm 10\%$ at $0.2\pi\lesssim q\lesssim 0.8\pi$.

Thus, our measurements present a detailed characterization of the
non-hydrodynamic part of the excitation spectrum in the 1D $S=1$
HAFM. We find that single-mode dispersion gradually crosses over
to a {\it narrow continuum} at $q\lesssim 0.5\pi$. The continuum
starts {\it below} the lowest possible energy of the two
non-interacting magnons, indicating their {\it attraction}.

We gratefully acknowledge discussions with
C.~Broholm, L.-P.~Regnault and A.~Zheludev, which inspired us
throughout this study. This work was carried out under Contract
DE-AC02-98CH10886, Division of Materials Sciences, US Department
of Energy. The work on SPINS was supported by NSF through
DMR-9986442.

\end{document}